\documentclass[11pt]{article}
\usepackage{amsmath,amssymb,bbold,epsfig}
\usepackage{amsfonts}

\newcommand{\be}{\begin{equation}}
\newcommand{\ee}{\end{equation}}
\newcommand{\ba}{\begin{eqnarray}}
\newcommand{\ea}{\end{eqnarray}}

\def\veps{\varepsilon}
\def\g{\gamma}
\def\p{\prime}

\begin{document}

\title{\Large \bf Synchrotron Radiation from Ultra-High energy protons
  and the Fermi observations of GRB 080916C}

\author{{Soebur Razzaque$^{1,2}$\thanks{email: \tt
srazzaque@ssd5.nrl.navy.mil}~, Charles D.\
Dermer$^{1}$\thanks{email: \tt
charles.dermer@nrl.navy.mil} ~and Justin D.\ Finke$^{1,2}$}
\\
\\
{\normalsize \it
$^{1}$Space Science Division, U.S. Naval
  Research Laboratory, Washington, DC 20375, USA}
\\ 
{\normalsize \it
$^{2}$National Research Council Research Associate}
}
\date{}

\maketitle

\begin{abstract} 
Fermi $\g$-ray telescope data of GRB 080916C with $\sim 10^{55}$ erg
in apparent isotropic $\g$-ray energy, show a several second delay
between the rise of 100 MeV -- GeV radiation compared with keV -- MeV
radiation. Here we show that synchrotron radiation from cosmic ray
protons accelerated in GRBs, delayed by the proton synchrotron cooling
timescale in a jet of magnetically-dominated shocked plasma moving at
highly relativistic speeds with bulk Lorentz factor $\Gamma \sim 500$, 
could explain this result. A second generation electron
synchrotron component from attenuated proton synchrotron radiation
makes enhanced soft X-ray to MeV $\g$-ray emission. Long GRBs with
narrow, energetic jets accelerating particles to ultra-high energies
could explain the Auger observations of UHE cosmic rays from sources
within 100 Mpc for nano-Gauss intergalactic magnetic fields.  The
total energy requirements in a proton synchrotron model are $\propto
\Gamma^{16/3}$. This model for GRB 080916C is only plausible 
if $\Gamma \lesssim 500$ and the jet opening angle is $\sim 1^\circ$.
\\
\\
{\it Subject headings:} gamma rays: bursts---gamma rays:
theory---radiation mechanisms: nonthermal
\end{abstract}

\section{Introduction}

An integrated fluence of $2.4\times 10^{-4}$~erg~cm$^{-2}$ was
measured from GRB 080916C with the Large Area Telescope (LAT) and
Gamma ray Burst Monitor (GBM) on the Fermi Gamma ray Space Telescope,
with one third of this energy in the LAT~\cite{grb080916c}. At a
redshift $z = 4.35\pm 0.15$~\cite{Greiner09}, GRB 080916C has the
largest apparent energy release yet observed from a GRB. A significant
$\simeq 4.5$~s delay between the onset of $>100$~MeV compared to the
$\sim 8$~keV -- 5~MeV radiation is found (the characteristic duration
of the GBM emission is $\approx 50$~s).

The spectrum of GRB 080916C was fit by the smoothly connected double
power-law Band function~\cite{Band93} to multi-GeV energies, though
with changing Band spectral parameters and peak photon energy in
different time intervals. The emergence of delayed spectral hardening
is represented by a Band beta spectral index changing from $\beta =
-2.6$ in the first 3.6 seconds following the GRB trigger to $\beta =
-2.2$ at later times~\cite{grb080916c}.  Here we show that a hard
spectral component arising from cosmic-ray proton synchrotron
radiation explains the delayed onset of the LAT emission. If GRBs
accelerate UHECRs, then the delayed onset of the LAT emission after
the GBM trigger should be a regular feature of GRB spectral evolution.

\section{Proton acceleration and radiation}

GRB blast wave calculations usually treat
electrons~\cite{Piran05,Meszaros06}, but protons and ions will also be
accelerated if they are present in the relativistic flows in
black-hole jet systems. Here we consider protons accelerated in GRB
blast waves to such energies that they can efficiently radiate hard
$\sim$ GeV -- TeV photons by the proton synchrotron
mechanism~\cite{bd98,tot98a,tot98b,ZM01,Huntsville08,wan09}.  The
highest energy photons are reprocessed by $\g\g\to e^+ e^-$
opacity~\cite{rmz04} to make an injection source of electrons and
positrons that cool by emitting $<$ GeV electron synchrotron
radiation. Two delays arise, the first from the time it takes to
accelerate protons to a saturation Lorentz factor where the
acceleration rate equals the synchrotron rate. A second delay arises
during which sufficient time passes to build up the spectrum of the
primary protons so that they become radiatively efficient in the LAT
band. Observations of GRB 080916C are consistent with the delayed
onset between GBM and LAT emission being caused by the second delay
where the evolving proton cooling synchrotron spectrum sweeps from
higher energies into the LAT waveband. This radiation is emitted from
a jet of magnetically-dominated shocked plasma with $\zeta_B >1$,
where $\zeta_B$ is the ratio of magnetic-field to proton/particle
energy density in the plasma.

The rapid variability, large apparent luminosity, and detection of
high-energy photons from GRBs can be understood if this radiation is
emitted from jetted plasma moving with bulk Lorentz factor $\Gamma\gg
1$ towards us.  Detection of 3 GeV and 13 GeV photons from GRB 080916C
suggests $\Gamma_3 = \Gamma/1000 \sim 1$~(Ref.~\cite{grb080916c} and
below, Section 3).  For variability times $t_v\sim 1$~s and
instantaneous energy fluxes $\Phi =
10^{-5}\Phi_{-5}$~erg~cm$^{-2}$s$^{-1}$, the internal radiation energy
density in the fluid is $u^\p_\g \approx 4\pi d_L^2\Phi /(4\pi
R^2c\Gamma^2) \approx (1+z)^2 d_L^2\Phi/(\Gamma^6c^3t_v^2)$, implying
a characteristic jet magnetic field of
\be
B^\p ({\rm kG}) \approx 2\frac{\sqrt{\zeta_B\rho_b\Phi_{-5}}}
{\Gamma_3^3t_v({\rm s})} \approx 2\zeta_B^{1/2}\rho_{b}^{1/2}
t_{v}^{-1/2}({\rm s}) E_1^{-1/2}(13\,{\rm GeV}),
\label{B_field}
\ee
where primes refer to the comoving frame, $\rho_{b}$ 
is the baryon-loading parameter giving the relative energy in
nonthermal protons compared to $\g$-rays.  The last relation in
equation~(\ref{B_field}) assumes $\Gamma \approx \Gamma_{\rm min}$
from the opacity condition $\tau_{\g\g} = 1$, which can be written as
\begin{equation}
\Gamma_{min} \cong \big[ {\sigma_{\rm T} d_L^2 (1+z)^2 
f_{\hat\epsilon}\epsilon_1\over 4t_v m_e c^4}\big]^{1/6}\;,
\label{Gmin}
\end{equation}
where $f_{\hat\epsilon}$ is the $\nu F_\nu$ flux at photon energy
$\hat \epsilon = 2\Gamma^2/[(1+z)^2 \epsilon_1]$ (in $m_ec^2$ units)
and $E_1 = m_ec^2 \epsilon_1$ is the highest energy photon.  Thus the
total jet energy is $\propto \zeta_B \rho_b\Phi_\g$.  A characteristic
field $B^\p \sim 10$ -- 100~kG is consistent with the absence of a
distinct self-Compton synchrotron component in GRB
080916C~\cite{grb080916c}, which happens when the magnetic energy
density is larger than the nonthermal electron energy density, or
$\zeta_B\rho_b \gg 1$.

In Fermi acceleration scenarios, protons gain energy on timescales
exceeding the Larmor timescale, implying an acceleration rate ${\dot
\gamma}^\p_{acc,p} = \phi^{-1}eB^\p/m_pc$, where $\phi^{-1} \ll 1$ is
the acceleration efficiency.  Equating the acceleration rate with the
synchrotron loss rate gives a saturation Lorentz factor for protons,
namely
\be
\gamma^\p_{sat,p} = \frac{m_p}{m_e} \left(
\frac{B_{cr}}{\phi B^\p} \right)^{1/2} \sqrt{\frac{9}{4\alpha_f}}
\approx \frac{2\times 10^8}{\sqrt{(\phi/10) B^\p_5}},
\label{p_sat}
\ee
where $\alpha_f = 1/137$ is the fine structure constant, $B_{cr} =
m_e^2c^3/e\hbar = 4.4\times 10^{13}$~G is the critical magnetic field,
and the mean magnetic field of the radiating region is $B^\p =
10^5B^\p_5$~G.  The observer measures a time
\be
t_{sat} = \frac{1+z}{\Gamma} \frac{m_p^2c}{m_e} \sqrt{ \frac{6\pi\phi}
{e\sigma_TB^{\p 3}} } \approx \frac{0.01\sqrt{\phi/10}} {\Gamma_3
B^{\p 3/2}_5}~{\rm s}
\label{t_sat}
\ee
for protons to reach $\gamma^\p_{sat,p}$.

The proton synchrotron saturation frequency (in $m_ec^2$ units),
corresponding to the proton synchrotron frequency of protons with
$\gamma^\p_{p} = \gamma^\p_{sat,p}$, is
\be
\veps_{sat,p} = \frac{\Gamma}{1+z}\veps^\p_{sat,p} =
\frac{\Gamma/\phi}{1+z} \frac{m_p}{m_e} \frac{27}{8\alpha_f} \approx
1.6\times 10^7\frac{\Gamma_3}{\phi/10}.
\label{e_sat_p}
\ee
This frequency corresponds to a photon energy of $\approx 8$~TeV for
GRB 080916C. Analogous to electron blast wave physics~\cite{spn98},
proton synchrotron losses make a cooling break at the proton cooling
Lorentz factor $\gamma^\p_{c,p} = \gamma^\p_{sat,p}(t_{sat}/t)$, which
is obtained by equating the synchrotron energy-loss timescale with the
comoving time $t^\p = \Gamma t/(1+z)$.  Consequently the proton
synchrotron cooling frequency for protons with $\gamma^\p_{p} \approx
\gamma^\p_{c,p}$ is $\veps_{c,p}(t) = \veps_{sat,p} (t_{sat}/t)^2$ for
our simplified model of continuous acceleration and uniform injection
of particles during the first $\sim$ 8~s to the observer.

Photons with energies above $\sim$ 1--10 GeV are strongly attenuated
through $\g\g\to e^+e^-$ processes in the source, inducing a
nonthermal $e^\pm$ injection that makes a second-generation electron
synchrotron component. Proton synchrotron photons radiated by protons
with $\gamma^\p_{p} = \gamma^\p_{sat,p}$ materialize into electrons
and positrons with Lorentz factor $\gamma^\p_e \approx
\veps^\p_{sat,p}/2$.  The electron synchrotron saturation frequency
from this second generation of synchrotron emission is observed at
\be
\veps_{sat,e} \approx \frac{3}{2} \frac{\Gamma}{1+z} \phi^{-2}
\frac{B^\p}{B_{cr}} \left( \frac{m_p}{m_e} \frac{27}{16\alpha_f}
\right)^2 \approx \frac{10^3\Gamma_3 B^\p_5}{(\phi/10)^2}.
\label{e_sat_e}
\ee
Therefore the second-generation synchrotron radiation cuts off above
$\approx 600$ MeV for GRB 080916C. This will be below the LAT
sensitivity because only the proton synchrotron radiation from protons
with $\gamma^\p_{p} \approx \gamma^\p_{sat,p}$ contributes to the
second-generation electrons and positrons that make emission at
$\veps\approx \veps_{sat,e}$ and the number of such protons is small
for a steep injection proton spectrum.

The direct proton synchrotron radiation from cooling protons equals
the electron synchrotron saturation frequency when $\veps_{c,p}(t) =
\veps_{sat,p} (t_{sat}/t)^2 = \veps_{sat,e}$, which is observed to
take place at time
\be
t_{cl} = t_{sat} \sqrt{\phi \frac{B_{cr}}{B^\p}\frac{m_e}{m_p}
\frac{64\alpha_f}{81}} = \frac{4}{3} \frac{1+z}{\Gamma}\phi
\frac{m_pcB_{cr}}{eB^{\p 2}} \sqrt{\frac{m_p}{m_e} } \;.
\label{t_close}
\ee
For GRB 080916C, $t_{cl}\approx 1.4{(\phi/10)}/{\Gamma_3^{-1} B^{\p
    -2}_5}$ s, which corresponds (depending precisely on
$\veps_{sat,e}$) to the time required for proton synchrotron radiation
to become strong in the LAT energy band. We propose this effect as the
reason for the delayed onset of the LAT emission.

In addition to the direct proton synchrotron radiation, pair
synchrotron radiation is formed from the internal attenuation
of proton synchrotron photons to make ultra-relativistic pairs.
If protons are accelerated with a number index $k$ (i.e., ${\dot
N}(\gamma) \propto \gamma^{-k}$), then the cooling spectrum breaks
from a proton distribution with index $k$ to a steeper one with index
$k+1$.  The proton synchrotron $\nu F_\nu$ flux, in the absence of
$\g\g$ opacity, is simply
\ba
\frac{f_{\veps}^{p,syn}}{f_{\veps_{sat,p}}^{p,syn}} = 
\begin{cases} ({\veps_{c,p}\over \veps_{sat,p}})^{\frac{2-k}{2}}
({\veps_{min,p}\over\veps_{c,p}})^{\frac{3-k}{2}}
({\veps\over\veps_{min,p}})^{\frac{4}{3}} , 
\veps < \veps_{min,p} \cr
({\veps_{c,p}\over\veps_{sat,p}})^{\frac{2-k}{2}}
({\veps\over\veps_{c,p}})^{\frac{3-k}{2}}, 
\veps_{min,p} < \veps < \veps_{c,p}(t) \cr
({\veps\over \veps_{sat,p}})^{\frac{2-k}{2}} ,
\veps_{c,p}(t) < \veps < \veps_{sat,p}
\end{cases}
\label{psyn_spectrum}
\ea
where $f_{\veps_{sat,p}}^{p,syn} = \Psi\rho_b\Phi_{\rm GBM}$,
$\gamma^\p_{min,p} \approx \Gamma_{rel}$, the relative Lorentz factor
of the relativistic wind in the jet and the shell material, and
\be
\Psi = \left(\frac{k-2}{2}\right) 
\frac{1- (\veps_{min,p}/\veps_{sat,p})^{(3-k)/2}}
{(\veps_{min,p}/\veps_{sat,p})^{(2-k)/2} -1} ;
\label{psyn_normalization}
\ee
$\Psi= [{\rm ln}(\gamma^\p_{sat,p}/\gamma^\p_{min,p})]^{-1} ~{\rm when}~k=2.$

The direct proton synchrotron flux from GRB 080916C is attenuated at
all energies from $\veps_{\g\g} \approx 1$ -- 10 GeV to $\veps \approx
\veps_{sat,p}$, equation~(\ref{e_sat_p}), to make a second-generation
electron injection spectrum with the same form as
equation~(\ref{psyn_spectrum}), though with subscript $p\to e$
spectral indices $(3-k)/2\to (3-k)/4$ and $(2-k)/2\to (2-k)/4$ and
$f_{\veps_{sat,e}}^{e,syn} = \frac{1}{2} f_{\veps_{sat,p}}^{p,syn}$.
The second-generation spectrum has a low-energy cut-off related to
$\veps_{\g\g}$, below which it receives no further injection pairs.
Note that $\veps_{c,e}(t) = \veps_{sat,e} (t_{sat}/t)^4$, where
$\veps_{sat,e}$ is given by equation~(\ref{e_sat_e}).

\begin{figure}
\vskip0.in
\center{\includegraphics[width=4.0in]{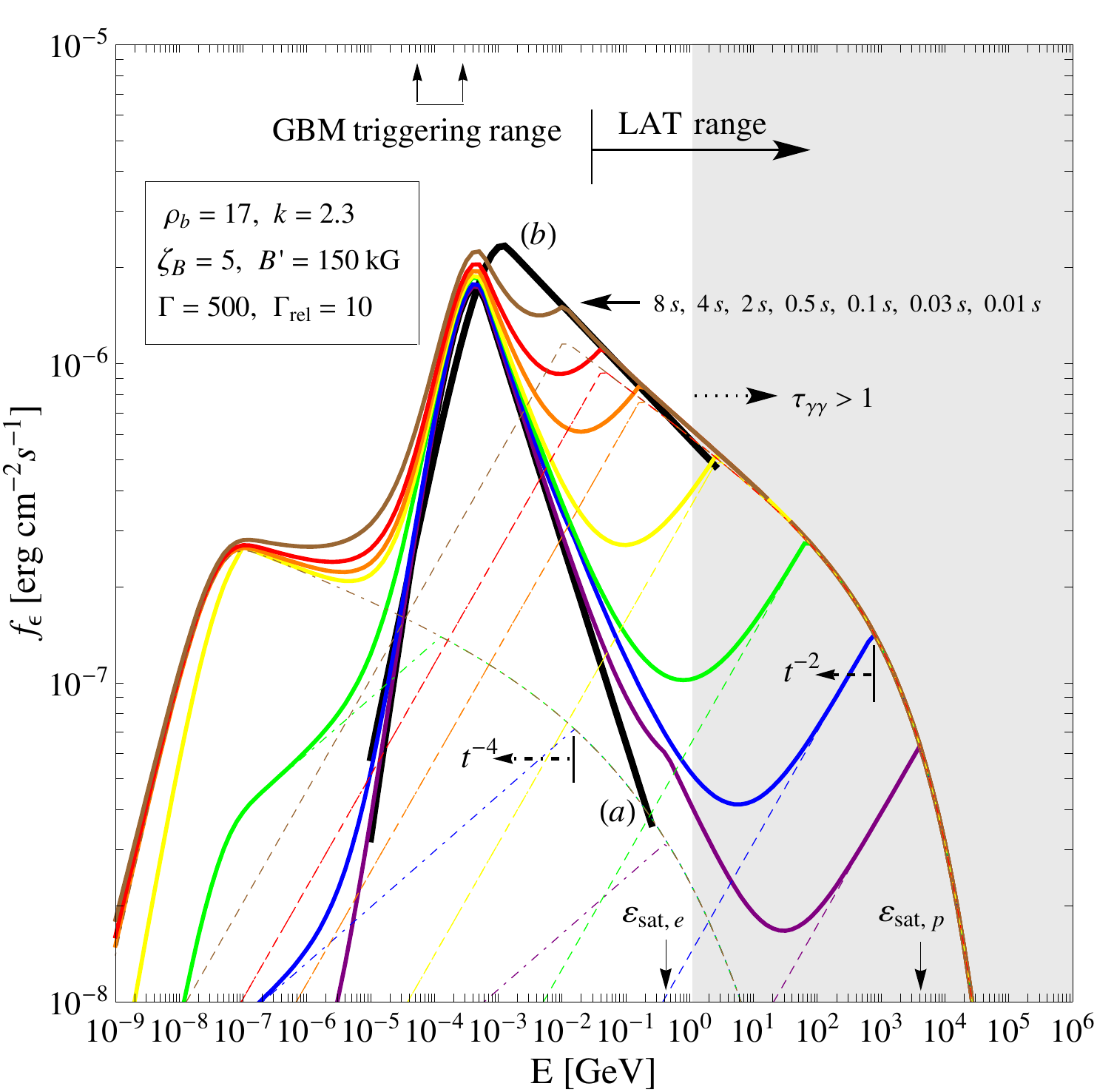}}
\vskip0.in
\caption[]{
Synchrotron model for the $\g$-ray spectrum of GRB 080916C. The $\nu
F_\nu$ flux in the GBM range is produced by primary nonthermal
electron synchrotron radiation. The heavy dark curves show the fitted
Band spectrum for intervals (a), from 0 to 3.6 s after the trigger,
and interval (b), from 3.6 s to 7.7 s. A strong proton synchrotron
component is formed at high energies and sweeps into the LAT band
after several seconds, thus making the time delay between the GBM and
LAT emissions.  A weak pair synchrotron $\g$-ray component is formed
in the GBM and LAT band from second-generation pair synchrotron
radiation formed by proton synchrotron photons that are attenuated by
$\g\g$ processes in the jet. Here we assume uniform injection over 8
s, and take $\phi =10$.
At a fixed photon energy below the cooling frequency, the $\nu F_\nu$
flux increases $\propto t$ for both the proton synchrotron and pair
synchrotron radiation. }  
\label{f1}
\end{figure}

Fig.~1 shows the evolving proton synchrotron spectrum for constant
injection with time using the above relations for parameters of GRB
080916C, with the Band function in interval (a), the first 3.6 s
following the trigger.  The Band function in the $\sim 10$~keV to
$\lesssim 100$~MeV range can be due to synchrotron radiation by
primary electrons, or thermal radiation from a jet photosphere.  For
the proton spectrum, we use $k=2.3$, $\rho_b = 17$, $\zeta_B = 5$, and
$\gamma^\p_{min,p}=10$.  In this simple picture of the GRB jet, the
protons are accelerated to the saturation Lorentz factor within
$\approx 0.01$~s, making a prompt second-generation electron
synchrotron spectrum too weak to be detected, followed after a few
seconds by strong direct proton synchrotron emission that sweeps into
the LAT band from high energies. The time delay in the high-energy
$\g$-ray flux occurs even for variable injection, provided it occurs
over $\sim 8$ s to the observer. Although the data for GRB 080916C are
consistent with a Band function, spectral analyses with the addition
of a proton synchrotron component will determine whether this more
complicated model is compatible with Fermi LAT GRB data.

Implicit in the treatment is that the wind impacts either a shell
ejected earlier by the central engine or some roughly uniform density
material that already existed in the surrounding medium.  The extent
of this material is $\langle r \rangle \approx c \Gamma^2 \Delta
t/(1+z)\approx 5\times 10^{16}\Gamma_3^2 (\Delta t/8$ s) cm.  The
magnetic field is thus assumed to be roughly constant over the first
$\sim 10$~s because of the constant density medium that is being swept
up at the shock.  This assumption is supported by the roughly constant
value of electron synchrotron energy flux $\Phi_{\rm GBM}$ measured
with the GBM, in the same time interval~\cite{grb080916c}.

\section{Total Energy of GRB 080916C}

The comoving synchrotron cooling timescale of an ion with atomic mass
$A$ and charge $Z$ is given by $t^\prime_{syn} = (A^3/Z^4) (m_p/m_e)^3
(6\pi m_ec/\sigma_{\rm T} B^2\gamma$).  The comoving peak synchrotron
photon energy is $\epsilon^\prime_{syn} =
(Z/A)(m_e/m_p)B\gamma^2/B_{cr}$. Equating the observer time $t_{syn} =
(1+z)t^\prime_{syn}/\Gamma$ for the emission to be radiated at
measured energy $m_ec^2\epsilon_{syn} = \Gamma\epsilon^\prime_{syn}/
(1+z)= 100 E_{100}$ MeV implies a comoving magnetic field $
B^\prime({\rm G}) \cong 2.0\times 10^5\;A^{5/3}Z^{-7/3} E_{100}^{-1/3}
t_{syn}^{-2/3}({\rm s})$ and an isotropic jet power, dominated by
magnetic-field energy, given by
\begin{equation}
L_B \cong \frac{ R^2 c\Gamma^2 B^{\prime 2}}{2} 
\cong {2\times 10^{58}\Gamma_3^{16/3}A^{10/3}t^{2/3}_{syn}({\rm s})
\over Z^{14/3}E_{100}^{2/3} } \;
{\rm erg~s}^{-1},
\label{LB}
\end{equation}
letting the blast-wave radius $R\cong \Gamma^2 c t/(1+z)$. For Fe ($A
= 56, Z = 26$), the power requirements are reduced by a factor
$\approx 0.17$. Here, however, we consider only proton synchrotron
radiation.

Eq. (\ref{LB}) shows that $L_B \propto \Gamma^{16/3}$ \cite{wan09}.
The absolute jet power varies as the square of jet opening angle
$\theta_j$.  In Ref.~\cite{grb080916c}, only the uncertainty in the
redshift was used to provide uncertainty in $\Gamma_{min}$ using
simple $\gamma\gamma$ opacity arguments, Furthermore, validity of the
cospatial assumption that the soft photons are found in the same
region as the hard photons was assumed in the calculation of
$\Gamma_{min}$.  We use a likelihood ratio test to calculate the
chance probability to detect a photon with energy greater than the
maximum measured photon energy in a given time bin.  For a $-2.2$
photon number spectrum, this test gives values of $\Gamma = (0.90,
0.82, 0.75)\Gamma_{min}$ for exponential escape and $\Gamma = (0.79,
0.58, 0.36)\Gamma_{min}$ for slab or spherical escape at the
(1$\sigma$, 2$\sigma$, 3$\sigma$) levels, respectively.

If Type Ib,c supernovae are the progenitors of long GRBs like GRB
080916C, then $\theta_j\gtrsim 0.8^\circ$~\cite{sod06}.  Taking a
conservative lower limit $\Gamma_{3} \cong 0.5$ gives the absolute
energy requirements for GRB 080916C of ${\cal E}\cong 3\times
10^{53}(\Gamma_3/0.5)^{16/3}(\theta_j/1~{\rm deg})^2 (t_{syn}/8~{\rm
  s})^{5/3}$ erg after integrating eq.\ (\ref{LB}) over time and
multiplying by a two-sided jet beaming factor $f_b \cong 1.5\times
10^{-4}(\theta_j/1~{\rm deg})^2$.  The rotational energy available in
a core collapse supernova could be as large as $\approx 5\times
10^{54}$ erg for a 10 $M_\odot$ core \cite{pac98}.

Although it appears to be a coincidence that $\Gamma\approx
\Gamma_{min}$ in our model, which is required because of excessive
powers when $\Gamma\gtrsim \Gamma_{min}$, more complicated geometries
might relax the bulk Lorentz factor requirement further.  If the inner
engine makes the prompt MeV radiation and residual shell collisions at
larger radii make LAT $\gamma$-ray photons, then $\Gamma$ could be as
low as $\sim 300$~\cite{li08}.  In this case, the absolute energy
release could be as low as $\approx 2\times 10^{52}(\theta_j/1~{\rm
  deg})^2$ erg. Even though the model in Ref.~\cite{li08} provides a
separate explanation for the delayed onset (and predicts that the
variability timescale of the $> 100$ MeV radiation is longer than the
keV/MeV radiation), lack of cospatiality does not guarantee delayed
onset. For example, photospheric emission with leptonic emission from
internal shells would neither be cospatial nor necessarily exhibit a
delayed onset.

The jet break time with apparent isotropic energy release $\approx
2\times 10^{57}$ ergs is $t_{br}\cong 0.3 (\theta_j/1~{\rm
  deg})^{16/3}n_0^{-1/3}$ d. The jet break would have taken place
before Swift slewed, at $\approx T_0 + 17.0$ hr, to GRB 080916C
\cite{hov08}, with a hard electron spectrum to explain the shallow
X-ray decay observed $\gtrsim 10^5$ s after the GRB \cite{str08}.

A $\approx 3$~s and $\approx 2$~s delayed onset of $\gtrsim 100$~MeV
emission detected in another long GRB 090902B~\cite{grb090902b} and in
the short GRB 081024B~\cite{grb081024b}, respectively, could be
explained with proton synchrotron radiation as we discussed here,
although lack of redshift measurement for GRB 081024B prohibits
calculation of the bulk Lorentz factor and the total $\g$-ray energy.
It is, however, interesting to note that Fermi-LAT-detected short GRBs
emit higher fluence in the LAT range than in the GBM
range~\cite{grb081024b,LD09} which could be due hadronic emission
processes.

\section{Ultrahigh-energy cosmic rays from GRBs}

GRBs have long been considered as candidate sources to accelerate
UHECRs~\cite{Waxman95,Vietri95}.  The energy of protons with
$\gamma^\p_{p} \approx \gamma^\p_{sat,p}$, if they were to escape from
the GRB blast wave, is $\approx 2\times
10^{20}\Gamma_3/\sqrt{(\phi/10)B^\p_5}$~eV, so GRB 080916C can in
principle accelerate UHECRs.  The Larmor timescale $t^\p_{esc}({\rm
  s}) \approx 0.2/\sqrt{(\phi/10)B^{\p 3}_5}$ at $\gamma^\p_{p}
\approx \gamma^\p_{sat,p}$ is much less than the light-crossing time
$t^\p_{lc} ({\rm s})\approx 200t_v({\rm s})$, so that escape depends
on transport in the jet plasma magnetic field impedes escape
\cite{Dermer07}. Photohadronic processes can assist escape by
converting protons to neutrons, but are more important when the
internal photon density is high, or when the variability timescale
$t_v$ is small~\cite{DA03}. When $t_v({\rm s}) < 0.01 (\Phi_{\rm
  GBM}/10^{-5}~{\rm erg~cm}^{-2} {\rm s}^{-1})/(\Gamma_3^4
\veps_{pk})$, then photopion losses are important in GRB 080916C (see
Ref.~\cite{asa09} for applications to GRB 090510). The shortest
variability timescale for GRB 080916C observed with INTEGRAL is 0.1
s~\cite{Greiner09}, so unless the corresponding size scale for the
radiating region was even shorter, we can neglect photohadronic
processes. But a shorter variability timescale would require a larger
bulk Lorentz factor $\Gamma$ in order to explain detection of
multi-GeV photons, in which case photohadronic efficiency is
reduced~\cite{Huntsville08}.

We estimate the rate of long-duration GRBs as energetic as GRB 080916C
within the $\approx 100$ Mpc clustering radius for UHECRs observed
with Auger~\cite{auger07,Auger09}.  For maximum total energy releases
of $\approx 10^{54}$~erg, the GRB 080916C jet opening angle $\theta_j
< 100/\Gamma = 0.1/\Gamma_3$. The inferred GRB rate of $\approx
2f_b$~Gpc$^{-3}$yr$^{-1}$ \cite{GPW05} at the typical redshift $z=$1
-- 2 is a factor $\approx 1/10$ smaller at $\approx 100 d_{100}$~Mpc
due to the star formation rate factor and a factor $f_b > 200$ larger
due to a beaming factor.  A $60E_{60}$~EeV UHECR is deflected by an
angle $\approx 4.4^\circ ZB_{nG} E_{60}^{-1} d_{100}^{1/2}
\lambda_1^{1/2}$ in intergalactic magnetic field with mean strength
$B_{nG}$~nG and coherence length of $\lambda_1$~Mpc. Deflection causes
dispersion in time of arrival of UHECRs~\cite{WC96} and increases the
apparent rate. The corresponding number of GRB sources within $\approx
100$ Mpc with jets pointing within $4^\circ$ of our line-of-sight is
$\approx 30 (f_b/200) B_{nG}^2 E_{60}^{-2} \lambda_1^{3/2}$.
Complications arising from nonuniform magnetic field geometries
(e.g.\ Refs.~\cite{Kashti08,Murase09}) can lead to different values in
the rate estimates, but still allow GRBs as UHECR hosts.  Thus if
typical long duration GRBs have a narrow, highly relativistic core
accelerating UHECRs, then long-duration GRBs could account for the
Auger events within the GZK radius.

\section{Discussion}

We have developed a hadronic model based on synchrotron radiation by
protons in a highly magnetized shocked plasma, to explain delayed
onset of high energy ($\gtrsim$ 100 MeV) emission observed with Fermi
from GRB 080916C.  This proton synchrotron spectral component, in
addition to the electron synchrotron and photospheric emission
\cite{pee07} that dominates in the keV -- MeV range, initially starts
at much higher energy and later sweeps into the Fermi LAT range, thus
causing a time delay in the prompt phase.  Our model is compatible
with an internal shocks scenario and could also be consistent with an
origin of the high-energy emission in GRB 080916C from an external
shock~\cite{KB09}.  The late time extended emission observed in GRB
080916C and other GRBs could be due to the slower cooling of the
protons in the early afterglow related to the external shock
\cite{bd98,ZM01}.

Our approach differs from the conclusions of \cite{ZP09} that the jet
outflow energy in GRB 080916C is dominated by Poynting flux rather
than particle energy.  Their conclusion depends on the assumption that
the engine radius $r_0\approx ct_v/(1+z)$, whereas $r_0$ might be much
smaller.  In our scenario, both the relativistic outflow and the shell
on which it impacts are particle energy dominated; only the shocked
fluid is highly magnetized.  The GBM emission could be nonthermal
electron synchrotron radiation that includes photospheric emission
made at smaller radii than the nonthermal synchrotron radiation.
Compton cooling of shocked electrons is suppressed in highly
magnetized shocked plasma, but could be important in a leptonic model
with external cocoon photons ~\cite{Toma09}. Our model for GRB 080916C
also applies to other Fermi LAT GRBs with comparable $\Gamma$ factors
and small beaming angles.

Besides GRBs, UHECRs might also be accelerated in other systems of
relativistic outflows, including low luminosity
GRBs~\cite{mur06,WRM07}, radio galaxies~\cite{Stanev05} and blazars.
UHECRs could be formed through neutron escape when photopion processes
are important, which will require IceCube neutrino
detections~\cite{icecube04} to establish.  In GRB 080916C, where
multi-GeV radiation is observed with Fermi, we have shown that
synchrotron radiation from ultra-high energy protons accelerated in
GRB jets explains the delay of the $> 100$ MeV LAT emission with
respect to the keV -- MeV GBM emission, and that long duration GRBs
are possible sites of UHECR acceleration.

\section*{Acknowledgements}

This work is dedicated to the memory of David L.\ Band.

We thank Armen Atoyan, Eduardo do Couto e Silva, Jonathan Granot,
Francesco Longo, Peter M\'esz\'aros, Nicola Omodei, Kenji Toma,
Xuefeng Wu, Ryo Yamazaki, and anonymous referees for comments.  This
work is supported by the Office of Naval Research and NASA.

\end{document}